\documentclass[9pt,twocolumn,twoside]{opticajnl} %

\journal{opticajournal} 

\setboolean{shortarticle}{true}

\usepackage{lineno}

\NewDocumentCommand{\suppress}{m}{} 

\usepackage{units}
\usepackage{caption}
\usepackage{subcaption}
\usepackage[colorlinks=true, allcolors=blue]{hyperref}

\dates{Accepted for publication in Optics Letters, includes a typo correction (in Eq.~7) w.r.t. the published version, compiled \today} 

\title{Shack-Hartmann wavefront sensor slope measurement algorithm robust to scintillation and sensor saturation}

\author[1]{Timothée VENE}
\author[1]{Laurent M. MUGNIER}
\author[1]{Aurélie MONTMERLE-BONNEFOIS}
\author[1]{Jean-Marc CONAN}

\affil[1]{DOTA, ONERA, Paris Saclay University, 92322 Châtillon, France}

\affil[*]{timothee.vene@onera.fr}

\begin{abstract}
High throughput optical communication links require the use of adaptive optics to compensate for atmospheric turbulence even in strong perturbation conditions and at low elevation. 
Considering the intensity fluctuations brought by scintillation, we present a new slope measurement algorithm for the Shack-Hartmann wavefront sensor, based on a Joint Maximum Likelihood approach. 
We show that this estimator is robust to sensor saturation and widens by more than two orders of magnitude the intensity range exploitable for reliable slope measurements compared to classical algorithms.
Consequently, by increasing the intensity assigned to the wavefront sensor, we also increase the number of valid measurements despite sensor saturation.
\end{abstract}

\setboolean{displaycopyright}{false} 

\begin{document}

\maketitle
\vspace{-1.8\baselineskip} 
\section{Introduction}
    \vspace{-.5\baselineskip}
    Optical communication links are expected to be the most promising solution to establish high throughput links in various cases, such as horizontal or satellite-to-ground communications.
    However, atmospheric turbulence represents a major problem for optical communication~\cite{andrews2002}. 
    Adaptive Optics systems are designed to compensate for the distortion of the beam wavefront induced by turbulence along the line of sight.
    However, in the strong perturbation regime, scintillation affects the measurement of phase deformations by the wavefront sensor~\cite{primmerman1995}.
    
    Considering a Shack-Hartmann wavefront sensor (SHWFS), scintillation within the pupil plane induces large intensity variations depending on the subaperture: some present a low Signal-to-Noise Ratio (SNR) while others have a high SNR but may feature saturated pixels, both leading to an increased measurement error.
    
    Therefore, it is crucial to use a reliable slope algorithm for a wide intensity dynamic range.
    Thomas \emph{et al.}~(2006)~\cite{thomas2006} and Nicolle \emph{et al.}~(2004)~\cite{nicolle2004} compared centroid estimators and the correlation to find an optimal estimator depending on the flux collected by a subaperture.
    Besides, Gao \emph{et al.}~(2024)~\cite{gao2024} presented an algorithm combining three estimators and selecting the right estimator depending on the SNR of the subaperture.
    Additionally, Plett~\emph{et al.} (1997)~\cite{plett1998measurement} suggested that it can be advantageous to use the SHWFS close to sensor saturation to decrease the number of low SNR subapertures, at the cost of biased measurements in saturated subapertures.

    Here we consider Maximum Likelihood (ML) estimators because of their appealing properties, thoroughly reviewed and applied to wavefront sensing in Barrett \emph{et al.}~(2007)~\cite{barrett2007maximum}, \emph{i.e.} they are optimal in the following sense: they are consistent, and asymptotically, they are unbiased and reach the Cramér-Rao lower bound. 
    One could contemplate adding prior information on the sought slopes like Sallberg \emph{et al.}~(1997)~\cite{sallberg1997maximum} did in low flux regime, thus switching to a Maximum A Posteriori (MAP) estimation. 
    We choose to defer the use of prior information to the wavefront reconstruction step, so as to keep the slopes estimation unbiased. 
    Indeed, an efficient wavefront reconstruction requires a good knowledge of the distribution of errors on the slopes, which is 
    biased when using a regularized slope estimation method such as MAP.
    
    In this letter, we use the ML framework to propose two new slope estimators maintaining high accuracy in all noise regimes and dealing explicitly with saturated pixels.

\vspace{-0.5\baselineskip} 
\section{Studied slope estimators} \label{Section: Model}
    \vspace{-.3\baselineskip}
    \subsection{Framework of the study}
        \vspace{-.2\baselineskip}
        We study the case of the image of a point source at the focal plane of a SHWFS subaperture.
        We suppose that the wavefront seen by the subaperture can be described by a combination of tip and tilt only.
        Therefore, the resulting image can be described as a shifted reference image of a point source.
        Besides, we assume that the subapertures are small enough so that the intensity distribution inside them is approximately uniform.
        Our image model at the pixel $(k,l)\in \mathbb{N}^2$  is thus:
        \begin{equation}
            \mathbf{i}(k,l) = \alpha ({r}*\delta_{xy})(k,l) +\mathbf{n}(k,l)=\alpha {r}(k-x,l-y) +\mathbf{n}(k,l) ,
        \end{equation}
        where $({r}*\delta_{xy})$  is a normalized reference image of a point source shifted at position $(x,y) \in \mathbb{R}^2$, $\alpha$ is the intensity in the subaperture varying drastically with time and between subapertures due to scintillation, and $\mathbf{n}$ is the noise affecting the image.
        We can condense our model as: 
        $\mathbf{i} = \alpha \left[r *\delta_{xy}\right]_{\mathrm{III}} +\mathbf{n}$,
        where $[.]_{\mathrm{III}}$ denotes the sampling operator. 
        We assume that any non-linearity of the image is calibrated and compensated for beforehand except for the saturation, which is studied later on.
        
        In order to estimate the shift $(x,y)$ of the image, we use a Maximum Likelihood (ML) approach which leads to the minimization of the neg-log likelihood:
        $J(x,y;\ \alpha) = -\ln(P(\mathbf{i}|x,y; \alpha ))$, where the flux $\alpha$ is first assumed known.
        
        We assume that the detector noise can be approximated with an homogeneous white Gaussian distribution of variance $\sigma_{det}^2$. 
        We approximate the photon noise by an inhomogeneous white Gaussian distribution of variance $\sigma_{phot}^2$ on each pixel.
        The resulting noise mixture is a Gaussian white distribution with a variance on each pixel equal to the sum of the noise variances~\cite{Mugnier-a-04}. 
        We can thus write $J$ as:
        \begin{equation} 
            J(x,y;\ \alpha) = \frac{1}{2} \sum_{k,l}\ \mathbf{w}(k,l)\  |\mathbf{i}(k,l)-\alpha r(k-x,l-y)|^2 +\mathrm{constant} .
            \label{neg-log likelihood +cst}
        \end{equation}
        with $\mathbf{w}(k,l)=\frac{1}{\sigma_{det}^2+\boldsymbol{\sigma}_{phot}^2(k,l)}$.
        Our estimator consists in searching the shift $(x,y)$ that minimizes Eq.~\ref{neg-log likelihood +cst}. With the above assumptions, this ML estimator is equivalent to a Weighted Least Square (WLS) approach.
        Eq.~\ref{neg-log likelihood +cst} can be developed as:
        \begin{equation}
            \begin{split}
            J(x,y;\ \alpha)  = \frac{1}{2} \sum_{k,l}\ \mathbf{w}\ \mathbf{i}^2(k,l)-2[\mathbf{w} \mathbf{i}](k,l)\ \alpha r(k-x,l-y) \\ 
            +\mathbf{w}(k,l)\ \alpha^2 r^2(k-x,l-y) +\mathrm{constant} ,
            \end{split}
            \label{neg-log likelihood +cst developped}
        \end{equation}
        We note that $\sum_{k,l}\ \mathbf{w}\ \mathbf{i}^2(k,l)$ does not depend on $(x,y)$, so we can discard it into the constant term. 
        We define a "correlation" evaluated at any non-integer shift $(x,y)$ by the following operation between a continuous function $f$ and a discrete array $\mathbf{v}$: 
        \begin{equation}
            \left[f \otimes \mathbf{v} \right]((x,y)) = \sum_{k,l} f(k-x,l-y) \mathbf{v}(k,l) .
            \label{def corr}
        \end{equation} 
        With this definition, we can simplify Eq.~\ref{neg-log likelihood +cst developped}:
        \begin{equation}
            J(x,y;\ \alpha)  = \frac{1}{2}\ \alpha^2 \left[r^2\otimes\mathbf{w}\right](x,y) - \alpha \left[r\otimes\mathbf{wi}\right](x,y)  +\mathrm{constant} .
            \label{neg-log likelihood +cst otimes}
        \end{equation}

    \vspace{-.5\baselineskip}
    \subsection{The maximum correlation estimator}\label{Subsection: correlation estimator}
        \vspace{-.2\baselineskip}
        If we consider that $\mathbf{w}$ is uniform, and by assuming small shifts and a large FoV, \emph{i.e.} negligible flux loss at the border of the image, the first term of Eq.~\ref{neg-log likelihood +cst otimes} does not depend on $(x,y)$.
        Then, we can take $\alpha$ out of the remaining term since it is only a constant multiplicative factor, and finally we find the following estimator:
        \begin{equation} 
            \begin{split}
            (\hat x, \hat y) &= \arg\max_{x,y}\left( \sum_{k,l}\mathbf{i}(k,l)\  \ r(k-x,l-y) \right)\\
            &= \arg\max_{x,y}\left( \left[r \otimes \mathbf{i}\right] (x,y)\right) .
            \end{split}
            \label{corr}
        \end{equation}
        Under these assumptions, the WLS estimator is actually a Least Square (LS) estimator, and it is equivalent to the maximization of the correlation, defined in Eq.~\ref{def corr}, between the image and the reference.
        Therefore, we can either minimize Eq.~\ref{neg-log likelihood +cst} or maximize Eq.~\ref{corr} to estimate $(x,y)$.
        In the rest of the paper, we will refer to this estimator as the correlation.

    \vspace{-.5\baselineskip}
    \subsection{The proposed estimators}\label{Subsection: The new estimator}
        \vspace{-.2\baselineskip}
        Due to the wide dynamic range induced by scintillation, some pixels of the image can be saturated.
        We use $\mathbf{w}$ to weigh the pixels depending on the uncertainty that affects them and propose setting the weight of saturated pixels to zero to discard these pixels from the estimation.
        Furthermore, in practice, $\alpha$ is unknown and we need to estimate it when $\mathbf{w}$ is not uniform.
        A simple approach would be to estimate $\alpha$ as: $\hat{\alpha}_s=\sum_{k,l} i(k,l)$.
        In the following, we refer to the combination of this approach and the minimization of Eq.~\ref{neg-log likelihood +cst otimes} as the Simple Weighted Least Square (SWLS) estimator:
        \begin{equation}
            \boxed{
                (\hat{x},\hat{y}) = \arg\min_{x,y} \left( \frac{1}{2}\ \hat{\alpha}_s^2 \left[r^2\otimes\mathbf{w}\right](x,y) \\
                - \hat{\alpha}_s \left[r\otimes\mathbf{wi}\right](x,y)\right).
            }
        \end{equation}
        However, when saturation of some pixels occurs, we can anticipate that this flux estimation may no longer be reliable. 
        For this reason, we choose to jointly estimate the flux $\alpha$ and the shifts $(x,y)$, within the same framework, and minimize the neg-log likelihood now denoted by $J'$ as a function of $x,y$ and $\alpha$: 
        $J'(x,y,\alpha) = -\ln(P(\mathbf{i}|x,y,\alpha ))$ . 
        
        Thus, we search for $\hat{\alpha}$ that solves $\frac{\partial J'}{\partial \alpha} = 0$ for any given value of $(x,y)$. 
        Because the criterion $J'$ of Eq.~\ref{neg-log likelihood +cst developped} is quadratic in $\alpha$, we easily obtain:
        \begin{equation} 
            \hat\alpha(x,y) = \frac{\sum_{k,l}(\mathbf{w\ i})(k,l)\ \ r(k-x,l-y)}{\sum_{k,l} \mathbf{w}(k,l)\ \ r^2(k-x,l-y)} = \frac{\left[r \otimes (\mathbf{w\ i})\right](x,y)}{\left[r^2 \otimes \mathbf{w}\right](x,y)} \ .
            \label{estim alpha}
        \end{equation}
        Then, we plug Eq.~\ref{estim alpha} into Eq.~\ref{neg-log likelihood +cst otimes} and finally, we obtain:
        \begin{equation}
            \boxed{(\hat x, \hat y)= \arg\max_{x,y}\left(\frac{\left[r \otimes (\mathbf{w\ i})\right]^2(x,y)}{\left[r^2\otimes \mathbf{w}\right](x,y)}\right)} .\label{estimateur double gaussien}
        \end{equation}
        We study also this estimator to measure slopes in the case of a large intensity dynamic range, and we call it the Joint Weighted Least Square (JWLS) estimator.
        We can either maximize the criterion of Eq.~\ref{estimateur double gaussien} or minimize the WLS criterion of Eq.~\ref{neg-log likelihood +cst} with $\alpha=\hat{\alpha}$ of Eq.~\ref{estim alpha}.
        
        In the following, we use the correlation as a comparison point to study the performance of both SWLS and JWLS estimators, because the correlation is a commonly used estimator in the literature~\cite{poyneer2005} and it is unbiased unlike centroid algorithms while presenting the same asymptotic behavior at high fluxes~\cite{thomas2006}. 
        This comparison also shows the interest of using a non-uniform $\mathbf{w}$.

\vspace{-0.5\baselineskip}         
\section{Method and simulation parameters}\label{section:method}
    \vspace{-.5\baselineskip}
    We consider a turbulence scenario representative of daytime dowlinks with a satellite at low elevations: $r_0=1-2\mathrm{cm}$ and $\sigma^2_\chi>0.5$. 
    The AO design features a $50\mathrm{cm}$ diameter telescope and a 25-by-25 subapertures SHWFS.
    In order to assess the performance of the SWLS and the JWLS estimators, we simulate occurrences of point source images at the focal plan of a subaperture in these conditions.
    We then compare the precision of our estimates with that of the correlation to quantify the improvement brought by our estimators.
    
    We first compute the shifted PSFs in the focal plane of a square lenslet.
    To do so, we apply a tilted wavefront at the pupil plane and then form the image at the focal plane of the lenslet.
    The shifts follow a Gaussian distribution centered at the center of the FoV and with a standard deviation of 0.3 pixel to be representative of the displacement of the PSF after an AO correction in strong perturbation regime.
    The PSFs are Shannon-sampled and computed on a field of 16x16 pixels before being cropped to 8x8 pixels to avoid periodization due to the use of FFT.
    We generate $N$ normalized images (typically $N=1000$) that we multiply by fluxes from $10^2$ photons to $10^7$ photons.
    {We consider such a large range because, in the conditions stated above, the flux collected by a subaperture spans over 3 orders of magnitude due to scintillation, around an average flux that depends on the incident power allocated to the SHWFS.}
    Then we add the noise: photon noise following a Poisson distribution; and readout noise following an homogeneous and centered Gaussian distribution with a standard deviation of 30 photoelectrons/pixel.
    Note that we could also take into account the photon noise caused by the background signal due to daytime operation by adding it to the RON value.
    Finally, the pixel values are capped above a certain value to simulate the sensor saturation.
    Here we consider that the saturation occurs at $3.5\ 10^4$ photoelectrons. 
    Both saturation level and readout noise values are typical orders of magnitude for a state-of-the-art InGaAs camera at high gain.
    To compute the weighting map $\mathbf{w}$, we estimate the noise variance as in~\cite{Mugnier-a-04}: the photon noise variance is estimated as the (noisy) image thresholded at zero and the detector noise variance is supposed to be calibrated beforehand.       

    There are two ways to perform the estimation for both SWLS and JWLS estimators: (1) search for the maximum of the criteria maps defined in Eq.~\ref{corr} and Eq.~\ref{estimateur double gaussien}. 
    We compute them on a grid with a 1-pixel pitch by means of FFTs and then interpolate them to reach sub-pixel precision;
    or (2) search for the minimum the WLS criterion of Eq.~\ref{neg-log likelihood +cst} after plugging the expression of $\hat{\alpha}$ into it.
    
    We choose to use the latter approach because, when using the first, we encountered a saturation of the measurement precision due to the precision limit of the maximum search algorithm, preventing the ultimate performance of the estimators, as already pointed out in the literature ~\cite{anugu2018}.
    Similarly, we minimize Eq.~\ref{neg-log likelihood +cst} with uniform $\mathrm{w}$
    for the correlation estimate.
    
    Furthermore, to avoid any local minimum, we initialize the minimization at the maximum pixel of the correlation computed on a grid with a 1-pixel pitch by means of FFTs.
    We choose this method because it is close to a first estimation of either studied estimator and it has a low computational cost.
    
    Concerning the correlation, according to Section~\ref{Section: Model}.\ref{Subsection: correlation estimator}, the equivalence between the maximization of the correlation and the minimization of the LS is valid for all values of $\alpha$.
    However, when numerically minimizing the LS criterion, we observed in practice that the estimation was better with an approximation of the intensity.
    Hence, we consider the correlation with the true value of the flux for a fair comparison with the SWLS and JWLS estimators.

    To evaluate the precision of both estimators, we calculate the Root Mean Square Error (RMSE) to the true position of the spot given the applied tilt: $RMSE = \langle\sqrt{(x-x_0)^2+(y-y_0)^2}\ \rangle_{N}$, where $(x_0,y_0)$ are the true coordinate of the shift and $\langle.\rangle_{N}$ is the mean operator on the number of occurrences N of both shifts $(x_0,y_0)$ and noise draws.
    This metric combines the bias and the variance of the estimation error.
    
\vspace{-0.5\baselineskip}    
\section{Results}
    \vspace{-.5\baselineskip}
    In this section, we first compare the two approaches to estimate the flux factor $\alpha$, and then we compare the performance of the JWLS, the SWLS, and the correlation, from a low signal case to a saturated signal case. 
    
\begin{figure}[!tb]
            \centering
            \captionsetup{justification=centering}
            \begin{subfigure}[c]{0.5\textwidth}
                \includegraphics[width=\linewidth]{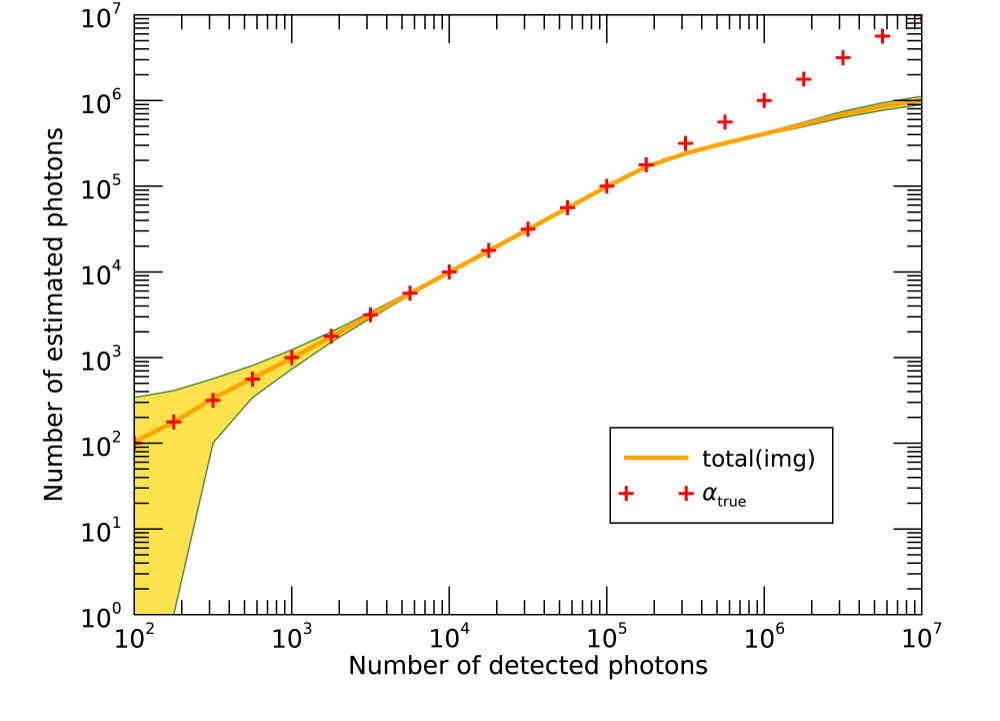}%
                \vspace{-1\baselineskip}
                \caption{Estimation of $\alpha$ using $\hat{\alpha}=\sum_{k,l} i(k,l)$.}
                \label{fig: alpha estimation A}
            \end{subfigure}

            \bigskip
            
            \begin{subfigure}[c]{0.5\textwidth}
                \includegraphics[width=\linewidth]{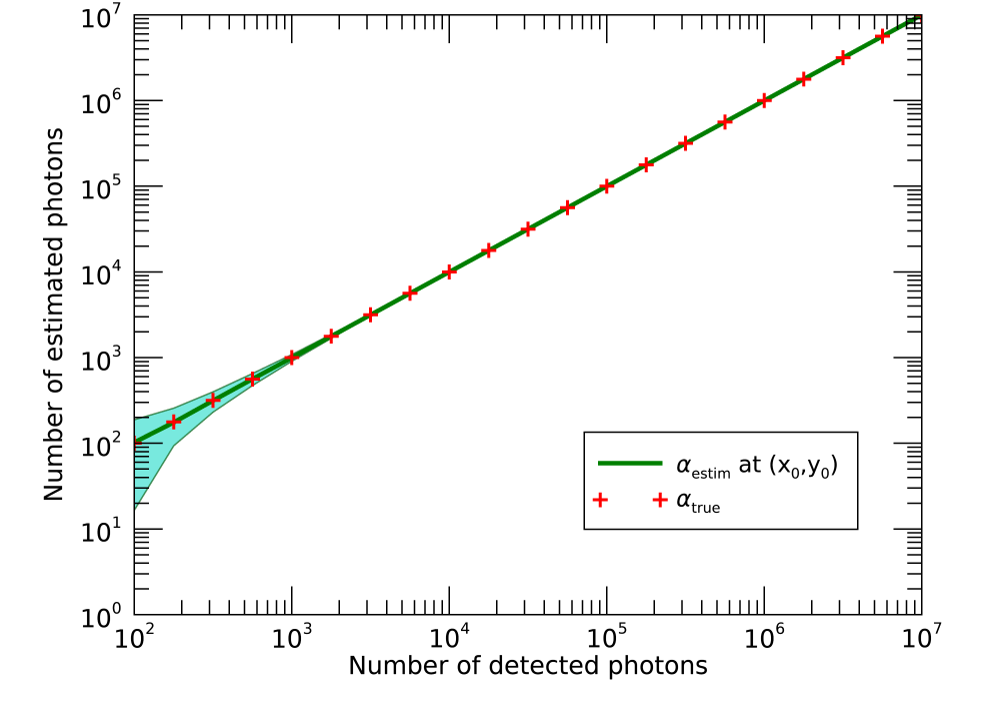}%
                 \vspace{-1\baselineskip}%
                \caption{Estimation of $\alpha$ using $\hat\alpha$ from Eq.~\ref{estim alpha} evaluated at $(x_0,y_0)$.}
                \label{fig: alpha estimation B}
            \end{subfigure}
            \vspace{-.5\baselineskip}
            \caption{Comparison of two approaches to estimate $\alpha$. The mean estimation in both cases is represented with a solid line, and the standard deviation is shown as a colored area. The pixels are saturated when their value exceeds $3.5\ 10^4$ photoelectrons.} 
            \label{fig: alpha estimation}
            \vspace{-1\baselineskip}
        \end{figure}

        In Fig.~\ref{fig: alpha estimation}, we show the reliability of the two approaches mentioned in Section~\ref{Section: Model}.\ref{Subsection: The new estimator} to estimate $\alpha_{true}$ (red crosses).
        
        Fig.~\ref{fig: alpha estimation A} depicts the simple and intuitive method, using $\hat{\alpha}_s$, the sum of the pixel values.
        The mean estimation (orange solid line) follows the true values of $\alpha$ as long as no pixel is saturated. 
        But, when saturation occurs, this method underestimates the flux because the pixels values are capped.
        Here, on average, the flux is mostly divided between four pixels, thus the effect of the saturation appears when the flux is higher than approximately four times $3.5\ 10^4$ photoelectrons.
        Furthermore, at low flux, as the pixels values are dominated by detector noise, summing their values results in a large variance of the estimation.
        
        Fig.~\ref{fig: alpha estimation B} corresponds to the second method, based on Eq.~\ref{estim alpha} and evaluated here at the true position of the spot $(x_0,y_0)$.
        We observe that the mean estimation (green solid line) is reliable and its variance is negligible even with saturated pixels. 
        This is expected since this method uses only unsaturated pixels.
        We can see on the figure that this methods fares also very well at low flux. 
        Indeed, in this case, the image is dominated by the detector noise and the weighting map $\mathbf{w}$ is almost uniform and equal to the inverse of the detector noise.
        The denominator of Eq.~\ref{estim alpha} is then uniform, and the reference in the numerator behaves as a filter that limits the number of noise dominated pixels taken into account in the estimation.
        This explains why the variance of this estimation is substantially smaller than the variance of the simple method.
        
        \begin{figure}[!htb]
        \centering
        \includegraphics[width=1\linewidth]{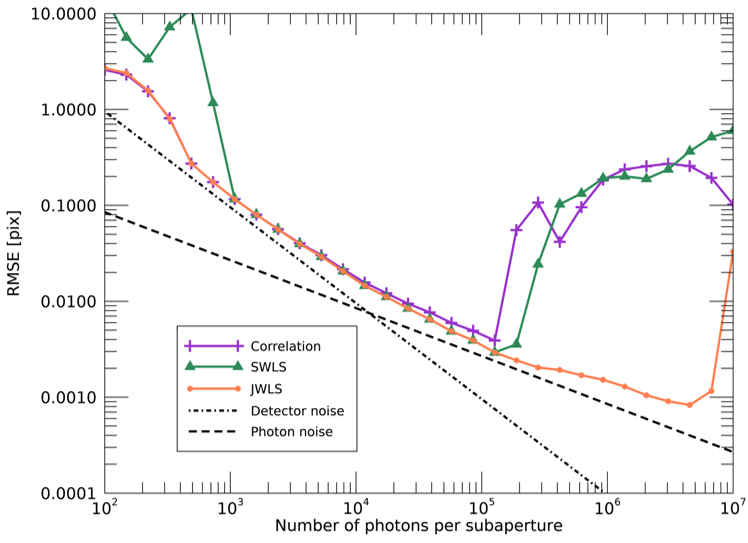}%
        \vspace{-1\baselineskip}
        \caption{Root Mean Square Error of slopes estimation as a function of the  number of photons in the image with sensor saturation at $3,5.10^4$ photoelectrons/pixel for: the correlation (purple line), the SWLS estimator (green line) and the JWLS estimator (orange line). The black dotted and dashed lines correspond to the correlation asymptotes in the detector noise and in the photon noise regimes respectively.}
        \label{fig: RMSE with saturation}
        \vspace{-1.4\baselineskip}
        \end{figure}
        In Fig.~\ref{fig: RMSE with saturation}, we show the RMSE of the slope estimate as a function of photon flux per subaperture with saturation.
        The JWLS estimator is represented by the orange curve, the SWLS estimator by the green curve and the correlation by the purple curve.
        We also show the asymptotes of the correlation error variance in detector noise and photon noise regimes from~\cite{thomas2006}.
        
        The first thing that stands out on the figure is that when increasing the flux above the saturation point at $10^5$ photoelectrons per subaperture, the correlation error starts to increase dramatically\footnote{We can also note the presence of two peculiar points on the purple curve at $4\ 10^5$ and at $10^7$ photons where the error is smaller while more pixels are saturated. This is due to the fact that, at these fluxes, the pattern created by saturated pixels is symmetrical around the true position of the spot resulting in a smaller bias.}, and at $2\ 10^5$ photoelectrons the same happens to the SWLS estimator error, whereas the error of the JWLS estimator still follows its asymptotic behavior up to $7\ 10^6$ photoelectrons, two orders of magnitude higher, when approximately 40\% of the pixels are saturated.
        This robustness to saturation is expected because we only consider pixels that are not saturated for the computation of the WLS estimators.
        By comparing the RMSE of both WLS estimators, we observe that the accuracy of the flux estimation is crucial, and that the better estimation of $\alpha$ yields the larger gain in term of operating domain.
        Moreover, what is not visible in Fig.~\ref{fig: RMSE with saturation} is that the saturation point of the JWLS depends on the size of the field of view (FoV) as more pixels remain for the estimation: the larger the FoV, the more robust the estimation.
        
        In the photon noise regime (from $10^4$ photoelectrons to the saturation point), the correlation error behaves as expected and follows its asymptote. And, since both estimations of the flux are valid in this regime, our slope estimators outperform the correlation through the weighing of the pixels.
        
        In the detector noise regime (below $10^4$ photoelectrons), the weighting map $\mathbf{w}$ is approximately uniform, hence the WLS is a simple LS equivalent to the correlation, if, as mentioned in Section~\ref{section:method}, the flux estimation is accurate.
        Since the joint ML approach to estimate $\alpha$ is precise enough, the JWLS estimator is performing as well as the correlation ; whereas the SWLS estimator RMSE increases drastically because its flux estimation is not sufficiently accurate.
        In this regime, given the distribution of the spots, it would be better to set the estimate to 0 rather than relying on any estimator by using a threshold on the intensity at $3\sigma_{det}$ for instance.
        In a real AO system this would result in a more stable closed loop.
        
        Furthermore, we have checked that our estimator is empirically unbiased (\emph{i.e} there is no offset between the true value and the mean estimation over noise occurrences and shifts), because the bias is approximately $10$ times smaller than the error standard deviations. 
        Concerning the initialization, it is crucial because the peak width of the JWLS criterion decreases as the flux increases. 
        To expand the domain of validity of the JWLS estimator, we can use a more precise initialization like a correlation computed on a 0.5-pixel pitch grid.
        However, we obtain a similar result when initializing with the JWLS instead of the correlation.
        
\vspace{-0.5\baselineskip}    
\section{Conclusion}
\vspace{-.5\baselineskip}
\enlargethispage{1\baselineskip}
    Considering scenarios of optical links in a strong perturbation regime, we proposed two new slope estimators that are precise on a wide SNR range: the SWLS and the JWLS estimator.
    
    We showed that both estimators are more robust to sensor saturation than the correlation, through the map $\mathbf{w}$ weighing the pixels depending on their variance and discarding saturated pixels from the estimation. 
    Additionally, due to a better flux estimation, the JWLS estimator proved to be the most robust to sensor saturation, and is the recommended estimator in this context.
    In the case of an AO system for an optical communications application, this algorithm enables a more flexible operating point since it is reliable on an intensity dynamic range two orders of magnitude larger than with a classical algorithm.
    For instance, to reduce the number of very low SNR subapertures, we can increase the share of the incident power on the SHWFS, even if it means saturating more pixels in high SNR subapertures.
    Moreover, the accurate estimation of the intensity in a subaperture despite noise and saturation will benefit other applications such as: the pre-compensation of a GEO-feeder link with a predictive approach~\cite{Lognone:23}, where spatio-temporal correlations of intensities are needed to control the AO loop; $C_n^2$ profile measurement on the line of sight~\cite{vedrenne2007c}, where both slopes and subapertures intensities must be estimated precisely; and also optical metrology. 
    
    In future works, we will study various implementations of the JWLS estimator in order to improve its cost efficiency.

\begin{backmatter}
\bmsection{Funding} The author would like to thank ONERA and AID for funding his PhD.
\vspace{-0.5\baselineskip}
\bmsection{Disclosures} The authors declare no conflict of interest.
\vspace{-0.5\baselineskip}
\end{backmatter}

\bibliography{opticsletter}

\end{document}